\documentclass[journal=jpclcd,manuscript=letter]{achemso}

\usepackage{xcolor}
\usepackage[utf8]{inputenc}
\usepackage{amsmath}
\usepackage{amssymb}
\usepackage{graphicx}
\usepackage{graphics}
\usepackage[T1]{fontenc}

\author{Stefano Mocatti}
\email{stefano.mocatti@unitn.it}
\affiliation{Department of Physics, University of Trento, Via Sommarive 14, 38123 Povo, Italy}
\author{Giovanni Marini}
\email{giovanni.marini-2@unitn.it}
\affiliation{Department of Physics, University of Trento, Via Sommarive 14, 38123 Povo, Italy}
\author{Matteo Calandra} 
\email{m.calandrabuonaura@unitn.it}
\affiliation{Department of Physics, University of Trento, Via Sommarive 14, 38123 Povo, Italy}

\title{Light-Induced Non-Thermal Phase Transition To The Topological Crystalline Insulator State In SnSe}

\begin{tocentry}
\includegraphics[scale=1.00]{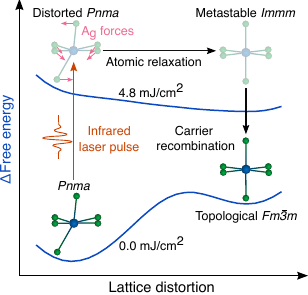}
\end{tocentry}
\begin{document}
\begin{abstract}
Femtosecond pulses have been used to reveal hidden broken symmetry states and induce
transitions to metastable states. However, these states
are mostly transient and disappear after laser removal. Photoinduced phase transitions towards
crystalline metastable states with a change of topological order are rare and difficult to predict and realize experimentally.
Here, by using constrained density functional perturbation
theory and accounting for light-induced quantum anharmonicity, we show that ultrafast lasers can
permanently transform the topologically-trivial orthorhombic structure of SnSe into the
topological crystalline insulating rocksalt phase via a first-order non-thermal phase transition.
We describe the reaction path and evaluate the critical fluence and the possible decay channels after photoexcitation.
Our simulations of the photoexcited structural
and vibrational properties are in excellent agreement with recent pump-probe data
in the intermediate fluence regime below the transition  with an error
on the curvature of the quantum free energy of the photoexcited state that is smaller than $2\%$.

\end{abstract}
\maketitle

The development of ultrafast laser light with femtosecond (fs) pulses has led to the possibility of inducing a substantial electron-hole population unbalance in semiconductors\cite{Maiuri2020}. After some tens of fs, this electron-hole plasma is well described by a two-chemical potential model, where both electrons and holes are characterized by a thermal distribution. Thus, the ions feel an out-of-equilibrium electronic population with a substantial occupation of conduction or antibonding states that can lead to structural phase transitions before electron-hole recombination takes place. In this scenario, ultrafast pulses can be used to overcome free energy barriers and synthesize crystal structures that cannot be reached by conventional thermodynamical paths. This kind of structural transformation is labeled {\it non-thermal}, to distinguish it from the much slower ones involved in conventional ({\it thermal}) material synthesis.
Experimental demonstrations of non-thermal phenomena induced by fs pulses are order-disorder phase transitions\cite{WallVO2}, charge density waves\cite{Kogar2019}, non-thermal melting of solids\cite{Solowski1995}, transient topological phase transitions\cite{Sie2019} and light-induced suppression of incipient ferroelectricity \cite{Jiang2016}. In all these cases, ultrafast light induces short-lived transient states. Much less common are light-induced \textit{non-thermal} permanent structural modifications.

\begin{figure*}[t!]
\centering
\includegraphics[width=1.00\textwidth]{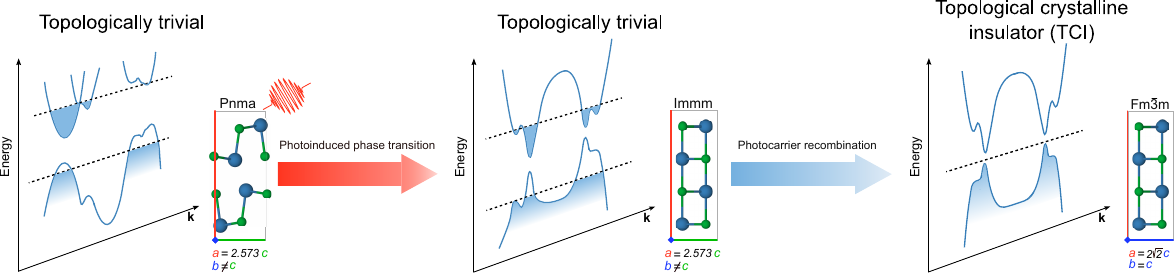}
\caption{Pictorial representation of the non-thermal pathway connecting the topologically trivial \textit{Pnma} structure with the TCI \textit{Fm}$\bar{3}$\textit{m} structure. For each phase, sketched crystal structures and band structures are represented. Fs pulses induce a first-order transformation toward a transient phase with {\it Immm} symmetry. This phase spontaneously decays into the \textit{Fm}$\bar{3}$\textit{m} structure after electron-hole recombination.}\label{fig1}
\end{figure*}

In this work, we show that non-thermal processes can be used to permanently stabilize topological crystalline insulating (TCI) phases\cite{Fu2011,Hsieh2012,Sun2013,Dziawa2012,Junzhang2017}. In this work, we will focus our attention on tin selenide (SnSe), a IV-VI p-type narrow gap semiconductor that has become popular due to its attracting thermoelectric properties\cite{Guo2015,Xie2021,Chandra2022,Zhao2014} ($zT=2.6$ at $T=923$ K).
At ambient conditions, tin selenide crystallizes in the orthorhombic \textit{Pnma} structure, sketched in Fig.~\ref{fig1}.
At $T\simeq 813$ K\cite{Chattopadhyay1986} or finite pressure \cite{Loa2015}, it undergoes a second order phase transition to an orthorhombic \textit{Cmcm} structure\cite{Loa2015,Ghosh2016,Zeitschrift1998,Aseginolaza2019}.
In SnSe the topological non-trivial state occurs neither in the {\it Pnma} phase, nor in the {\it Cmcm} phase, but in a metastable rocksalt structure, which cannot be reached via a thermal transition but it can only be synthesized in thin films via epitaxial growth techniques on a cubic substrate\cite{Mariano1967}. 

Here we design a different approach to obtain the topological crystalline phase of SnSe (see Fig.~\ref{fig1}), namely we consider the effect of ultrafast pulses on the topologically trivial $Pnma$ phase, which is close to a band inversion\cite{Hong2019}.
By laser pumping with a near-infrared pulse (1.55 eV) and monitoring the time evolution with time-resolved Raman and  X-ray diffraction, it was recently shown that structural modifications occur in SnSe, signalled by $A_{g}$ modes softening and fluence-dependent atomic displacements\cite{Huang2022}, interpreted as the precursor of a symmetrization towards a different orthorhombic structure with {\it Immm} symmetry. 
However, no transition to this new crystal phase was detected and first principles simulations were unable to reproduce the observed structural distortion.

In this paper, we investigate the non-thermal structural transformations of the {\it Pnma} structure after irradiation with fs pulses by combining constrained density functional perturbation theory (c-DFPT)\cite{Marini2021} and Stochastic Self-Consistent Harmonic Approximation (SSCHA)\cite{Monacelli2021},
accounting for quantum anharmonicity in the presence of an electron-hole plasma for the first time. Further technical details are provided in the Supporting Information (SI), which includes Refs. \citenum{Giannozzi2009,Giannozzi2017,Hamann2013,Perdew_1996,Grimme2010,Monkhorst1976,Sundaram2002,MarzariVanderbilt1999,RevModPhys.73.515,Kokalj2003,Momma2011,Bansal2016,Li2015,Gong2020,Shi2015,Blochl1994,Nikolic1978,Huang2023}.
We identify the non-thermal pathway (see Fig.~\ref{fig1}) and the critical fluence for the structural transition from the ground state {\it Pnma}  to the transient 
{\it Immm} phase. Our calculated structural distortions and softenings of the $A_{g}$ modes along the reaction path are in excellent agreement with experimental data \cite{Huang2022}. Most importantly, we show that the transient {\it Immm} phase spontaneously decays into the TCI rocksalt \textit{Fm}$\bar{3}$\textit{m}  SnSe structure after electron-hole recombination and that the structural transformation is permanent in virtue of the free-energy barrier between the rocksalt and the $Pnma$ phase.

\begin{figure*}[t!]
\centering
\includegraphics[width=1.00\textwidth]{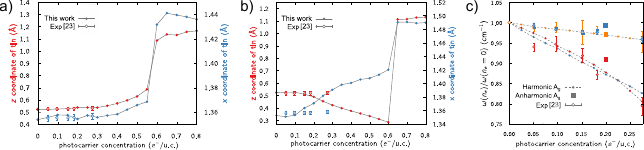}
\caption{Panels a,b: Tin Wyckoff positions $x$ and $z$ as functions of photocarrier concentration for fixed (a) and variable (b) volume crystal structure optimization. The red and blue dots label the $z$ and $x$ coordinates of tin, respectively. Panel c: Normalized phonon frequencies at $\Gamma$ for the three relevant $A_g$ modes versus PC. The red, blue and orange dots stand for the $A_{g,1},A_{g,2}$ and $A_{g,4}$ modes, respectively. The theoretical harmonic, anharmonic and experimental values of $\omega_0$ are reported in the SI. The experimental data are from Ref.~\citenum{Huang2022}. }\label{fig2}
\end{figure*}

In Fig.~\ref{fig2}(a),(b) we display the optimized tin Wyckoff $x$ and $z$ coordinates as functions of the photocarrier concentration (PC), $n_e$, expressed as the number of photoexcited electrons per unit cell (u.c.). Our results are compared
with time resolved diffraction data from Ref. \citenum{Huang2022} (see also SI, Sec.S3) measured in the first $5$ ps after illumination. 
After photoexcitation, both the internal equilibrium positions and the lattice parameters can change. However, the timescale for the two phenomena is generally different\cite{Reis2006}. To unambiguously disentangle cell deformation and internal displacements at fixed cell, we perform structural optimization at fixed cell, Fig.~\ref{fig2}(a), and at variable cell, Fig.~\ref{fig2}(b) in the presence of an electron-hole plasma (the procedure regarding fluence / PC mapping is reported in the SI). 

Our calculation at fixed cell is in excellent agreement with time-resolved X-ray diffraction data in the first 5 ps, while that at variable cell substantially deviates. This confirms that in the first ps after irradiation the atoms are displaced at fixed cell. Previous calculations from Ref. \citenum{Huang2022}, see Fig.S5,  obtained Sn displacements one order of magnitude larger than the experimental ones. Conversely, we report an excellent agreement between our c-DFT calculations and experimental data within our framework\cite{Marini2021}. The stark disagreement among the calculation of Ref. \citenum{Huang2022} and experiments arises because in 
Ref. \citenum{Huang2022} the
electron and hole occupations are not self-consistently relaxed. This procedure does not lead to a correctly thermalized quasi-equilibrium Fermi-Dirac distribution. An explanation of the main differences between the approach of Ref.\citenum{Huang2022} and the c-DFT approach of Refs. \citenum{Tangney2002} and \citenum{Marini2021} used in this work is reported in Sec.S3.1 in the SI.

The discontinuity in the $z_{Sn}$ and $x_{Sn}$ in Figs.~\ref{fig2}(a),(b) at $n_e^c\approx 0.6 \ e^-/$u.c. signals the occurrence of a first-order phase transition.
The phase transition can be easily identified by noting that for $n_e\ge 0.6 \ e^-/$u.c. the Wyckoff positions of tin correspond to that of the \textit{Immm} structure,
where they are fixed by symmetry. 
Thus, we predict the phase transition from \textit{Pnma} to \textit{Immm} to occur at a value of $n_e$ that is
approximately a factor of two larger than the highest photocarrier concentration measured in  Ref.~\citenum{Huang2022}. 

Additional validation of our findings arises from the $A_{g}$ harmonic and anharmonic phonon frequencies calculation at fixed cell. The results are shown in Fig.~\ref{fig2}(c) where they are compared with the measured frequencies of oscillation of the Bragg peaks in the first ps after pumping as a function of $n_e$\cite{Huang2022}. We plot the value of the harmonic $A_{g}$ phonon frequencies (full circles) at a given photocarrier concentration  (i.e. $\omega(n_e)$) divided by the harmonic phonon frequency in the ground state (i.e. $\omega_0=\omega(n_e=0)$). The softening of the harmonic modes $A_{g,1}$ and $A_{g,4}$ induced by the photoexcitation is in excellent agreement with experimental data. Concerning the $A_{g,2}$ mode, c-DFPT overestimates the softening induced by the electron-hole plasma within the harmonic approximation. A possible reason for this discrepancy is the presence of a strong anharmonic renormalization for the $A_{g,2}$ mode. Thus, we calculated the anharmonic phonon frequencies in the absence of photocarriers and for $n_e=0.2 \ e^-$/u.c. at $T=0$ K. Our results are depicted in Fig.~\ref{fig2}(c), where the values of the normalized anharmonic phonon frequencies ($\omega(n_{e})/\omega_0$) are represented as filled squares. The anharmonic corrections to the phonon frequencies of the $A_{g,1}$ and $A_{g,4}$ modes are mild and do not change the overall trend obtained at the harmonic level. On the contrary, the $A_{g,2}$ mode is substantially affected by anharmonicity, resulting in an improved agreement with the experimental data. The maximum relative error on the predicted quantum phonon frequency softening is roughly 2\%.

The $A_{g,2}$ mode plays a crucial role in the $Pnma \to Immm$ phase transition. The observed strong anharmonic renormalization of the $A_{g,2}$ mode indicates that the Born-Oppenheimer free energy surface surrounding the minimum energy state corresponding to the $Pnma$ phase is becoming progressively more anharmonic along the path of the transition.

\begin{figure*}[t!]
\centering
\includegraphics[width=0.95\textwidth]{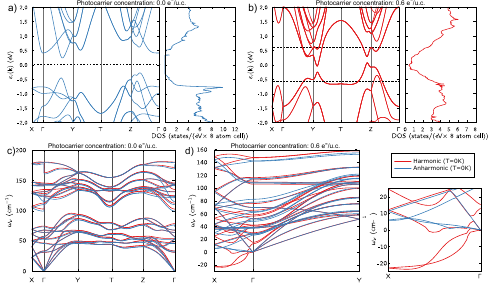}
\caption{Panels a,b: Ground state electronic structure of the {\it Pnma} phase (a) and of the {\it Immm} transient phase at $n_{e}=0.6 \ e^-/$u.c. (b). The Fermi level in (a) and the holes and electrons Fermi levels in (b) are depicted as dashed lines. Panels c,d: harmonic and anharmonic phonon spectra for the {\it Pnma} phase at $n_e=0.0 \ e^-/$u.c. (c) and for the transient {\it Immm} phase at $n_e=0.6 \ e^-/u.c.$ (d). Both plots are at $T=0$ K. The inset shows the removal of the dynamical instability by quantum anharmonic effects.}\label{fig3}
\end{figure*}

The crucial role of quantum anharmonicity becomes even more evident if the electronic structures together with the harmonic and quantum anharmonic dispersions of the {\it Pnma} and {\it Immm} phase are considered, in Fig.~\ref{fig3}). The insulating ground state (Fig.~\ref{fig3}(a)) displays a finite electronic gap ($\approx 0.52$ eV), and dynamically stable harmonic phonons  (Fig.~\ref{fig3}(c)). The quantum anharmonic corrections on the phonon spectrum are essentially negligible. 
On the contrary, the light-induced \textit{Immm} phase in the presence of an electron-hole plasma at $n_e=0.6 \ e^-/$u.c. has a metallic electronic structure with electron and hole Fermi surfaces located close to the high-symmetry Y and Z points,  (Fig.~\ref{fig3}(b)). These Fermi surfaces are nested (see SI, Sec.~S3) and trigger the emergence of a Peierls instability at the harmonic level signalled by imaginary phonons along the $\Gamma-X$ direction at a wave-vector compatible with the nesting condition (Fig.~\ref{fig3}(d)).
Structural minimization shows the emergence of a $ 2\times 1\times 1$ one-dimensional chain-like charge density wave
with an energy gain of $\approx$1.5 meV/atom with respect to the undistorted structure (see SI). 
When quantum-anharmonic corrections are included within the SSCHA at $n_e=0.6 \ e^-/$u.c., we find that the instability is removed and a sharp one-dimensional Kohn-anomaly appears (Fig.~\ref{fig3}(d)). Thus, light-induced quantum anharmonicity stabilizes the {\it Immm} phase in the transient state at $n_e=0.6 \ e^-/$u.c..

The critical PC of $n_e=0.6 \ e^-/$u.c., corresponding to $\approx 4.8$ mJ/cm$^2$, is achievable in ultrafast experiments similar or larger values have already been achieved in narrow gap semiconductors without inducing significant damage to the sample\cite{Hu2015,Cavalleri2001,wall_disordering_dimers}. 

\begin{figure*}[t!]
\centering
\includegraphics[width=0.95\textwidth]{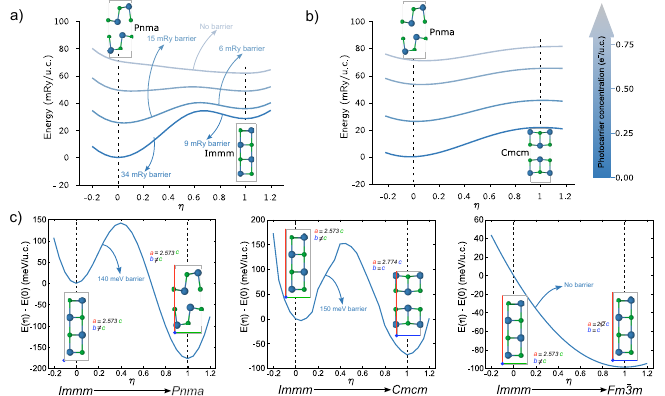}
\caption{Panels a,b: Total energy curves along the $Pnma\to Immm$ (a) and $Pnma \to Cmcm$ (b) reaction paths: $\eta=0$ corresponds to the $Pnma$ phase while $\eta=1$ to the $Immm$ (a) or $Cmcm$ (b) phase. Panel c: possible decay channels for the $Immm$ phase: $Immm\to Pnma$, $Immm\to Cmcm$ and $Immm\to Fm\bar{3}m$. We recall that 1 mRy/u.c. corresponds to 40 K while 1 meV/u.c. corresponds to 3 K.}\label{fig4}
\end{figure*}   

Having demonstrated the accuracy of our approach to describe the structural evolution after photoexcitation, we now try to understand more in detail the reaction path and the nature of this transition. In Fig.~\ref{fig4}, we display the energy along the paths relative to the (a) \textit{Pnma}$\to$\textit{Immm} and (b) \textit{Pnma}$\to$\textit{Cmcm} transitions, for a few values of $n_e$. The path is parametrized by the reaction coordinate $\eta$, where $\eta=0$ stands for the \textit{Pnma} structure while $\eta=1$ represents either the \textit{Immm} or \textit{Cmcm} structure. 

Considering the $n_e=0.0 \ e^-/$u.c. case, both the reactions toward the \textit{Immm} and \textit{Cmcm} present a large kinetic barrier. As the PC is increased, the \textit{Pnma}$\to$\textit{Immm} barrier is gradually suppressed and becomes zero for $n_{e} \simeq n_e^c$. Since the lowest energy configuration corresponds to the \textit{Immm} phase for $n_e > n_e^c$, the \textit{Pnma}$\to$\textit{Immm} reaction becomes spontaneous.  Conversely, the \textit{Pnma}$\to$\textit{Cmcm} barrier remains finite for every value of PC.

The question arises if the structural transformation towards the \textit{Immm} is permanent, i.e. if the \textit{Immm} phase remains stable at longer times after carrier recombination has taken place. To correctly describe the slow structural dynamics, one must also include volume relaxation effects.

To this aim, we consider variable-volume reaction paths in the absence of photoexcitation starting from the  $Immm$ structure, namely $Immm\to Pnma$, $Immm\to Cmcm$ and $Immm\to Fm\bar{3}m$. The initial $Immm$ structure corresponds to the photo-induced transient phase while the final structures are obtained through variable volume optimization with zero PC. Along the reactions, both the internal coordinates and the structural parameters vary. The results of our calculations are shown in Fig.~\ref{fig4}(c). Both the transformations $Immm\to Pnma$ and $Immm\to Cmcm$ present large energy barriers, thus are not spontaneous. Conversely, the $Immm\to Fm\bar{3}m$ reaction does not have a barrier and can occur spontaneously. Hence, once the transient $Immm$ phase has been stabilized, electron-hole recombination takes place and the system decays into the TCI $Fm\bar{3}m$ phase.
 
In addition, we stress that a large free energy barrier, amounting to 15 mRy/u.c., exists between the $Fm\bar{3}m$ and the \textit{Pnma} (see Fig. S6). Hence, the topological rocksalt phase can survive thermal fluctuations corresponding to $\approx 600 \ K$ before decaying into the \textit{Pnma} phase.
This finding demonstrates the occurrence of a non-thermal path stabilizing the SnSe rocksalt structure and provides a non-thermal synthesis mechanism for the rocksalt TCI phase.

We stress the fundamental role played by light-induced symmetrization. The TCI phase of rocksalt-SnSe is protected by a combination of time-reversal and C$_4$ symmetry, the latter being absent both in the $Pnma$ and $Immm$ phases. Crucially, we showed that laser irradiation favors the $Pnma \rightarrow Immm$ symmetrization, allowing the crystal to access a metastable region of the phase diagram, in close proximity to the \textit{Fm}$\bar{3}$\textit{m} structure, which is successively stabilized after the laser removal, finally restoring the cubic C$_4$ symmetry necessary to protect the topological crystalline order\cite{Fu2011}.

In conclusion, we have shown that ultrafast pulses can permanently transform the topologically-trivial $Pnma$ phase of SnSe into the TCI rocksalt phase. The mechanism is non-thermal and does not require epitaxial growth on particular substrates. This is one of the rare cases when ultrafast pulses change the topological properties of the material. We identified the transition path and evaluated its critical fluence. 
A strong validation for the accuracy and predictivity of our theoretical framework is the excellent agreement of our quantum anharmonic calculations in the photoexcited regime with recent pump-probe X-ray free electron-laser measurements in the low fluence regime below the transition. 

Finally, we would like to point out that our findings demonstrate that light can be used to reshape the free-energy landscape allowing to access otherwise unreachable regions of the phase diagram. This general result is relevant for the exploration of new phases in a broad class of materials, including monochalcogenides\cite{Behnia_Science}, highly relevant for energy applications, insulating/semiconducting 2D materials with strong spin-orbit coupling and, in general, all semiconducting materials in the proximity of structural instability. 

Funded by the European Union (ERC, DELIGHT, 101052708). Views and opinions expressed are however those of the authors only and do not necessarily reflect those of the European Union or the European Research Council. Neither the European Union nor the granting authority can be held responsible for them.
We acknowledge the CINECA award under the ISCRA initiative, for the availability of high-performance computing resources and support. We acknowledge PRACE for awarding us access to Joliot-Curie at GENCI@CEA, France (project file number 2021240020). 

\begin{suppinfo}
Description of the computational methods, crystal structures and comparison of the two different cDFT approaches. 
\end{suppinfo}

\bibliography{bibliography}
\end{document}